# Scalable Multi-Agent Lab Framework for Lab Optimization

A. Gilad Kusne[1,2,*] (ORCID: 0000-0001-8904-2087)  Austin McDannald[1] (0000-0002-3767-926X)
1. Materials Measurement Science Division, National Institute of Standards and Technology, Gaithersburg MD 20899
2. Materials Science & Engineering Dept, University of Maryland, College Park MD 20742
* Corresponding Author and Lead Contact: aaron.kusne@nist.gov

## Summary

Autonomous materials research systems allow scientists to fail smarter, learn faster, and spend less resources in their studies. As these systems grow in number, capability, and complexity, a new challenge arises – how will they work together across large facilities? We explore one solution – a multi-agent laboratory-control framework. The framework is demonstrated with autonomous material science labs in mind, where information from diverse research campaigns can be combined to address scientific questions. The framework can: 1) account for realistic resource limits, e.g., equipment use, 2) allow for research-campaign-running machine-learning agents with diverse learning capabilities and goals, 3) facilitate multi-agent collaborations and teams. The MULTI-agent auTonomous fAcilities Scalable frameworK (MULTITASK) makes possible facility-wide simulations, including agent-instrument and agent-agent interactions. Through modularity, real-world facilities can come on-line in phases, with simulated instruments gradually replaced by real-world instruments. We hope MULTITASK opens new areas of study in large-scale-autonomous and semi-autonomous research campaigns and facilities.

## Introduction

Autonomous research systems[1] are a paradigm shifting method for performing systematic scientific research. For these systems, a machine learning agent is placed in control of automated equipment to control experiment design, execution, and analysis, with the system learning and evolving with each new datum in an evolving environment. A machine learning function is often used as a surrogate for the unknown relationship between experiment inputs and outputs, mapping the complex and vast search space. Data collected from experiments are preprocessed and then used to improve the surrogate function's predictive power. Experiment design and selection is guided by active learning[2] – the machine learning field of optimal experiment design. Subsequent experiments are selected to provide maximal knowledge to the users.

With such machine learning pipelines, order-of-magnitude improvements in research efficiencies are possible[1]. Even greater efficiencies are possible by incorporating prior knowledge of the research challenge - such as prior theory and exogenous experimental and simulation data - into the pipeline[3]. These efficiency gains are exemplified by recent successes of autonomous physical science in optimizing mechanical structure[4], material processing[5], device operation protocol[6], and mixture rates of liquid-synthesized materials[7–9]. Examples in solid-state material exploration and discovery include mapping the composition-phase and composition-temperature-phase[10–12] relationships as well as the first autonomous discovery of a best-in-class material – a phase change memory material[11].

As autonomous systems grow in number, capability, and complexity, a new challenge arises – how will these systems work together? Autonomous control can be extended throughout labs, research and development facilities, and warehouses to achieve many orders-of-magnitude improvements in efficiencies. The benefits may be significant. For example, greater facility efficiencies may be gained by addressing multiple correlated challenges in concert, with each challenge benefiting from knowledge gained for the others. Such autonomous systems would free technical experts to focus on posing questions and gaining insight from results. As the scale of these systems grows, so too does the scope of the questions that can be addressed.

Within a facility-scale architecture, there are different ways to organize the interactions of these autonomous systems, in particular their machine learning agents – each with its own goals. The simplest option for the agent architecture is perhaps autonomous control by a central machine learning agent that collects all data and makes all decisions. This architecture has its benefits, including reducing the need for redundancies in data storage and computational resources by centralizing both. Another architecture is one with a central decision-making agent combined with distributed learning-only agents, exemplified by federated learning[13]. This architecture reduces the network burden of transmitting all data to a central site. A third architecture of multiple independent agents[14] working in concert, has its own advantages. These advantages mirror those of having multiple human experts working together to solve large challenges. For instance, agents of differing algorithmic bias (similar to individual human bias due to education and training) can validate each other's results.

How the goals of each agent relates to those of the others and to the overall facility goal can be informative of the optimal architecture. For example, consider the collaboration between two agents working on disparate tasks. Collaboration between the agents may provide research acceleration when the data from one task is informative of the other (e.g., phase maps and material properties[11]). Thus, improving overall facility performance. However, if data sharing is extraneous, it may serve only to increase the computational cost for data analysis tasks. In the latter case, sharing only the learning from each agent (and not the data) may be preferable. Similarly, distributing decision making may be computationally preferable for agents with disparate goals.

The laboratory framework presented herein is flexible enough to allow for different agent-control architectures including centralized agent, federated learning agents, and multiple independent agents – i.e., different modalities for a laboratory management system. Dynamic architectures are also possible, optimizing to match time-varying challenges. The framework is dubbed the MULTI-agent auTonomous fAcilities - a Scalable frameworK aka MULTITASK. This framework can demonstrate clear technical benefits. For example, facilities may have systems from which collected data is large (volume, velocity, etc.), resulting in network bandwidth issues when centralizing data. Distributed agents can take advantage of edge computing and data storage to overcome this challenge by performing local experiment design, data collection and reduction, and data analysis[15]. Also, specialized agents can be paired with instruments that require specialized machine learning tools for data analysis, resource optimization, and decision making.

MULTITASK also offers the flexibility of modularity, allowing real-world autonomous spaces to come on-line in stages, with simulated instruments gradually replaced by real-world instruments. As facilities change in size and scope, and even grow to contain distributed instruments (e.g., across states or countries), a networked, multi-agent framework can scale to demand, varying the number and type of agents. If two disparate facilities overlap in some instruments, successful agents from one facility can be copied to the other. Additionally, local agents can focus on local challenges and collaborate with agents working on larger scale challenges, such as agents working to optimize both local and shared resources. Such lab modularity formalizes the collaborative interactions which can be especially important when dealing with data privacy and proprietary information. For example, a localized agent could serve as a gatekeeper, providing information of internal signal trends while maintaining intellectual property protections through data privacy, federated learning, and other techniques.

We demonstrate MULTITASK in the context of an autonomous experimental material science laboratory. While the framework is general and can be easily extended to other fields or applications, experimental material science is a particularly informative example since many of the research efforts require input from multiple instruments. For example, to explore a class of materials for a particular application 1) a sample or set of samples must first be synthesized, 2) the structure and microstructure of those samples must be characterized by a suite of instruments, and 3) the relevant properties of the samples must be measured by a different suite of instruments before 4) the sample performance can be evaluated and understood.

MULTITASK's integrated multi-agent system[14] includes agents with autonomy in sensing, decision making, communicating, and (direct) acting. Agents are heterogenous in learning, goals, and decision making and they collaborate in either leaderless or leader-follow relationships. The collection of agent-agent and agent-instrument relationships form the network topology, with associated practical limits defined by available instruments. Successful agent collaborations require a set of agreement parameters. These are again lab-based, including a description of the target material system space as defined by lab capabilities, pertinent materials and instrument physics, and instrument resource limits and queues. Agent-run research campaigns are composed of discrete events that incorporate lab-based sample synthesis, transportation, measurement, and sharing delays. Research campaigns in a fully automated lab will not require agent (or autonomous system) physical mobility as only samples and resources move. However, labs undergoing automation can benefit from mobile autonomous systems[8] with associated agents.

The research lab provides further unique challenges for a multi-agent system. Agents follow the experimental process of sample synthesis, characterization, and analysis. Agents may face hysteretic or quantum mechanical challenges, e.g., the act of sensing may alter the target domain. Research campaigns are exploratory (heavily unsupervised) without clear end states. As a result, active learning methods are preferable over the more common use of multi-agent reinforcement learning. Furthermore, consensus between agents is achieved through multi-modal hypothesis testing – i.e., agents collaborate by playing each other's devil's advocate (which differs from typical game-type multi-agent collaborations). Additionally, competing agents may avoid each other in the target domain, seeking orthogonal paths toward the same goal (which differs from typical game-type multi-agent competitions). Through these interactions, hypotheses evolve from surrogate models to heuristics to potential physical models.

In this work, MULTITASK manages a simulated materials research lab with the goal of autonomous materials optimization. Past work into agent-based machine learning for the materials sciences includes three studies. The first study utilizes data processing agents and learning agents for identifying the composition-phase map from previously performed X-ray diffraction experiments[16]. Prior physics knowledge of phase mapping is encoded as a set of constraints, with different sets of learning agents able to apply different sets of constraints. One set of learning agents works together to identify viable phase map descriptions for small regions of the composition space. These results are unified by another set of learning agents to generate multiple viable solutions. The second study uses agents to identify a stable, optimal material by running density functional theory simulations[17]. In this study, a set of identical agents are instantiated, each runs a fully independent campaign, and results are combined at the end. The third study describes a framework where again independent agents (called 'orchestrators') run independent research campaigns[18]. For all these studies, 'agents' refer to independent learners, optimizers, or algorithmic units with predefined interactions. As agents do not have autonomy of sensing, action, or communication, these solutions are best described as distributed problem solving or parallel artificial intelligence rather than multi-agent systems[14]. To the authors' knowledge, there have been no investigations of multi-agent systems used in the context of autonomous facilities optimization.

The contributions of this work are:
- A unifying framework for an autonomous research lab, and a basis for similar facilities. The framework allows for real-world modules to come online unit by unit, replacing simulations and digital twins.
- A visual language for describing autonomous laboratory architecture. This combines visualization of multi-agent networks[19] with visualization of laboratory instruments and processes.
- Demonstrations of autonomous Bayesian learning agents controlling a simulated materials research laboratory toward the two goals of materials system exploration and optimization. Different lab architectures are compared.
- An algorithm for coregionalization of classification and regression developed for learning the composition-structure-property relationship of solid-state materials.

In the next section Framework Description, we describe the working parts of the framework, present example code for instantiating an autonomous facility, and demonstrate a visual language for communicating different instantiations. In Demonstration, we describe a materials optimization challenge and MULTITASK results. For more information about implementation, see the Methods section. We conclude with the Discussion section. We begin with a brief description of materials optimization and discovery.

## Background: Materials Discovery

The discovery of novel solid-state materials is key to the success of numerous next-generation technologies such as quantum computing, carbon capture, and low-cost medical imaging. These materials must possess advanced properties selected for their technology applications. To find these advanced materials, researchers utilize a fundamental relationship between how a material is made and its resulting structure and properties - the synthesis-structure-property relationship[20] (SSPR). For example, a material's properties are dependent on its elemental composition (e.g., iron, copper, etc.) and its phase – a description of the atomic organization within the material.

A useful tool in the search for advanced materials is the 'composition-phase map' which maps unique elemental compositions (and potentially other properties such as the temperature at which the material was synthesized) to the material's phase[20]. The phase of a material is determined through characterization techniques including X-ray diffraction, Raman spectroscopy, and transmission electron microscopy. Phase maps are segmented into regions separated by boundaries, known as phase regions and phase boundaries, respectively. Optimal materials of certain properties tend to occur within specific phase regions (e.g., magnetism and superconductivity) or along phase boundaries (e.g., caloric-cooling materials). Additionally, the functional behavior of material properties from phase region to phase region can vary, indicating a piece-wise nature dependent on structure. Thus, materials researchers identify and utilize composition-phase maps to guide their search for advanced materials.

## Framework Description

A selection of 'objects' (for background on object-oriented programming, see: [21]) can be instantiated, consisting of: agents, resources such as instruments, individual samples (i.e., items) used for the machine learning campaigns, and repositories for either physical samples or data. Physical instruments include those used for sample synthesis, processing, and measurement. Instruments also include computational tools used to generate sample simulation data (not used in the demonstration). Physical and computational instruments are represented as shared resources that produce or consume countable objects. For example, a synthesis instrument produces physical samples and devices while a computational instrument produces data for simulated samples and devices.

To instantiate an agent, synthesis instrument, measurement instrument, or other object just requires calling the class, i.e,. `agent()`, `instr_synthesis()`, `instr_measure()`. In Python, list comprehension can be used to instantiate a list of these objects with a call such as: `agents = [ agent(index, *properties) for index, properties in agent_descriptions ]`. These objects can be combined into facility units with multiple agents, shared instruments, and shared sample and data repositories.

### Agents

Agents consist of four basic properties. *Internal representation (IR)*: An internal representation of the world. This includes a perception of the materials search space, the tools and samples available, past collected data, and the other agents. *Goals*: Each agent has a set of goals with an associated set of active learning acquisition functions, which quantify the utility of future experiments. Agents can work together in a group by combining their acquisition functions (i.e., goals), to identify the next set of experiments that benefit the group as a whole. *Machine learning (ML)*: They have machine learning capabilities for data analysis, prediction, and decision making (i.e., active learning). Each agent can have its own unique ML capabilities. In principle these capabilities can be any sort of algorithm for data analysis and prediction, be that deep learning models, Bayesian inference models, physics-based analytical models, etc.

In the present work, agents employ Bayesian models as they are particularly well suited for active learning. When sharing data between agents, agents employ a coregionalization learning algorithm to exploit shared trends across the data sources (See Methods Section). *Communication*: Agents use functions for requesting, sending, and receiving data to share knowledge between individual agents, groups of agents, or with the full agent community through a central data repository. The four basic capabilities of IR, ML, goals, and communication are demonstrated through agent-agent collaboration for accelerated research campaigns.

### Instruments

Instruments can perform certain operations with associated capacity and delays. An example set of instruments for materials or device research would consist of sample synthesis, processing, property simulation, and characterization instruments. For instance, a synthesis instrument can be defined so that it makes 2 material samples at a time, with a synthesis time of 10 minutes. Agents place requests to instruments to perform desired operations (e.g., a synthesis instrument to make a sample, a measurement instrument to measure a sample) and these requests are put into operation queues for the instrument. In performing their operations, instruments can draw on or create limited, consumable resources. For example, a sample synthesis instrument may consume 1 silicon wafer to produce one material sample.

### Repositories

Two types of repositories exist – ones that store physical samples and ones that store data. Each type has a dedicated management system for agent interaction. *Sample repository*: Once a sample is synthesized, it is transferred to a sample repository to await a request to the repository management system for the sample to be processed or measured. When the sample is unused, it is returned to the sample repository. The sample-repository relationship is analogous to that of a book and library. The management system allows agents to identify which samples are in the repository and which are being lent out. *Data repositories*: These serve as central data storage facilities, where agents can share collected data including measurement data as well as data analysis, prediction, and decision-making acquisition function data. Management provides agent read and write access to databases.

### Visualizing the Network

A diagrammatic language can facilitate network description and comparisons through displaying the lab network infrastructure. Here we use shaped icons to indicate the physical and computing instruments (rectangle), repositories (hexagon), agents (oval), and humans (cut-off oval). Different forms of agent-agent and agent-instrument interactions can be indicated through the graph connections. Here data sharing is shown with a graph edge (connection between objects, i.e., nodes). The additional icon ► indicates that this data sharing path also permits sharing data on acquisition function which impacts decision making. The direction of the icon indicates the hierarchy of decision making with { Leader ► Follower}. If both directions are shown, both agents share their acquisition functions and impact the other's decision making. Edges marked with ■ indicate physical sample sharing between objects. An example is shown in Figure 1.

## Results

The coupled challenges of materials exploration and optimization present an excellent opportunity to demonstrate the multi-agent approach. More particularly, the fundamental dual challenge of exploring the synthesis-structure part of the SSPR through phase mapping, and the challenge of identifying materials with optimal functional property from the synthesis-property part of the SSRP ('synthesis' here includes composition). Through shared trends, knowledge of one can provide knowledge of the other, as demonstrated by recent autonomous materials discovery informed by the SSRP[11]. Agent collaboration can thus result in accelerated goal achievement.

For this work, individual agents either seek to maximize knowledge of the synthesis-structure relationship or seek the synthesis composition that maximizes the target property. The former agents are labeled 'PM' for phase map and the latter are labeled 'FP' for functional property. In collaboration, the agents combine data of composition, structure, and property to learn the SSPR relationship and thus improve analysis and prediction of their individual target objectives. By further combining their acquisition functions, agents can balance their goals with that of the community, selecting subsequent investigations that benefit overall community knowledge.

The materials data challenge is drawn from the perovskite oxide material system as characterized by [22] for piezoelectric response and structure. Data for samples near the $Bi_{1-x}Sm_xFeO_3$ edge of this composition space are shown in Figure 2. For this system, composition, lattice structure, and piezoelectric properties are strongly linked. Figure 2(A1) shows the relationship between composition and the piezoelectric coefficient $d_{33}$ (pm/V) which has its maximum near the boundary between phase regions (2) and (3). This is an easy maximization challenge as the extrema is characterized by a peak profile that stretches the full composition domain. Experiments performed on either side can use simple gradient ascent to find the maximum. A synthetic and more difficult challenge is presented in Figure 2(A2). Here the maximum is characterized by a highly local peak profile in phase region (3) and phase regions (1) and (2) are described by their own broad peak profiles and local maximums. Additionally, for samples near the $Bi(Fe_{1-y}Sm_y)O_3$ composition binary, structure is measured using Raman spectra, with example spectra shown for phase region (1) through (3) in Figure 2B along with the constituent lattice space groups.

An agent requests a composition to investigate, the sample is synthesized if needed, and the agent then requests for the sample to be characterized for either $d_{33}$ or Raman spectra based on their agent type (PM or FP). Synthesis, characterization, and sample requests from the sample repository have associated delays and queues. Agents, instruments, and data repositories are instantiated with facility units composed of the simulated physical portion diagramed in Figure 3A and a selection of the software portion given by either: Fig 3B) fully independent agents (named: Independent), Fig 3C) agents that share sample data (named: Data Sharing), or Fig 3D) agents that share sample data and where the PM agent shares acquisition function results with the FP agent for joint decision making (named: Data Sharing and Joint Decision Making). Agents with access to disparate types of data utilize a multi-modal learning algorithm as described in the Methods. The 'M' in the top right corner of each of the 'plates' in Fig 3 indicates that these objects are reproduced M times, as in the plate notation of probabilistic graphical models[23] (similarly with 'N'). For this demonstration, M=N=2. All objects associated with the physical lab share the same sample repository and all objects for the software portion of the lab share the same data repository.

Diagrams of agent performance for each of the 3 facility types is shown in Figure 4A1 and 4A2 for the two challenges. For the simpler challenge, the pair of independent synthesis-property agents (black line) identify the best material in 8 experiment design, execution, and analysis iterations. The more advanced agents show a dramatic improvement at lower iterations, with the agents sharing only data coming close to the best material in 7 iterations and the agents with joint decision making coming close to the best material in only 4 iterations. For the significantly more complex Challenge 2, knowledge of the phase boundary locations provides a significant boost to both types of data sharing agents bringing them close to the optimal material within 6 iterations while one of the non-sharing agents got stuck in a local optimum.

Figure 4B1 and 2 show the 95 % confidence intervals for the performance mean for each architecture and for each challenge over ten runs. Data sharing provides a significant performance boost for both challenges and joint decision making provides a moderate performance boost for challenge 2. Figure 4C displays an example of acquisition function combination. Figure 4D displays example functional property predictions from FP agents after 10 iterations under the various architectures shown in Figure 3. In this case, one of the non-data sharing agents (shown in Figure 4(D2)) identifies the optimal material despite operating under the incorrect assumption that the functional property has a uniform behavior (covariance length scale) across the entire composition space. However, this result is less robust than either of the Data Sharing architectures as evidenced for Challenge 2 (shown in Figure 4(D1)), where the mean performance for the independent architecture fails to find the optimal material after 10 iterations. Comparatively, both the FP agents with Data Sharing (seen in Figure 4(D3)) and Data Sharing and Joint Decision Making (seen in Figure 4(D4)) find the optimum within the 10 iterations. Furthermore, agents with Data Sharing both collect many data points near the optimum within the 10 iterations, suggesting the robustness of this behavior.

## Discussion

This preliminary work demonstrates the different performance achievable using diverse autonomous laboratory architectures. For these simple challenges, agent-agent data sharing and joint decision making performed better than only data sharing, which itself performed better than independent agents. The presented framework can be extended for diverse research possibilities. For example, one can investigate: the impact of agent heterogeneity in learning and decision-making capabilities; the relationship between agent-facility architecture and challenge type and difficulty; as well as agent-agent interactions. This latter possibility includes investigating agents with internal representations of other agents (e.g., expected beliefs and actions), adversarial agents capable of introducing poisoned data or models to throw off their competitors, and collaborative agents working together to improve the certainty of their individual results (i.e., cooperative validation).

Furthermore, just as multi-agent frameworks have been used to study markets[24] and political systems[25], the MULTI-TASK autonomous laboratory framework provides an opportunity to model the interactions of physical scientists. The framework can be used to evaluate the impact of varying scientist capabilities and behaviors, their facilities, and their scientific challenges. MULTITASK can then be used to investigate how scientist-scientist interactions, resource and instrument limitations, and others scientific factors impact learning, decision making, and discovery.

## Experimental Procedures

**Resource Availability**
**Lead Contact:** A. Gilad Kusne, aaron.kusne@nist.gov
**Materials Availability:** No materials were produced in this work.

**Code and Data Availability**: Both code and data are available on Github (https://github.com/KusneNIST/MULTI-TASK_Matter) and are part of the upcoming NIST Hermes library to be found at https://pages.nist.gov/remi/

## Agents: Independent Learning

*Independent phase mapping* is performed by first clustering the materials $x_s$ with collected Raman data using Spectral Clustering with a cosine measure applied to the Raman data (here intensities measured at different Raman shift values form individual vectors. The cosine measure is applied between vectors to define dissimilarity). The result is an approximate phase region labels $y_s$ for each sample. A full Bayesian extrapolation of phase region knowledge is then performed using Bayesian inference. For Bayesian inference step $i$ two change points values $c_i = \{c_{i,1}, c_{i,2}\}$ are sampled uniformly over the composition domain. These change points are then used to define a categorical distribution $M_s$ for the structure measured materials $x_s$ with the boundary between categories defined by the change points, using one-hot encoding. The sum log likelihood $L_s$ of the observed phase region labels, given the categorical distribution is then computed. The set of Bayesian inference samples $C = \{c_i\}$ approximate the posterior probability for $c$. The samples are then averaged to form a posterior probability for class membership over the full range $X$. This Bayesian inference operation is performed using the Pyro*[26] package and allows for probabilistic structure labels as inputs.

$$\underline{\text{Bayesian inference step } i:}$$
$$c_i \sim Uniform(X)$$
$$M_{i,s} = membership(c_i, x_s)$$
$$L_{i,s} = \sum_j \ln[Categorical(y_s(j)|M_{i,s})]$$

$$\underline{\text{Bayesian analysis}}$$
$$P_M = \text{mean}[membership(c_i, X)]$$

$P_M$ the posterior for membership probability will tend to have greater uncertainty at the unknown changepoints.

*Independent functional property regression* is performed using the Matern52 kernel over the composition domain with the standard Gaussian process regression function using the GPFlow package[27] for materials $x_f$ with measured functional properties $y_f$. The output is a Bayesian Gaussian process with an estimate of the most likely $y(x)$ paired with quantified uncertainty.

## Agents: Multimodal Learning

A custom Bayesian inference-based, full Bayesian coregionalization function is used to combine the tasks of phase mapping and materials property regression. First samples with Raman are clustered as with the *independent phase mapping*. The cluster labels define potential phase regions.

Within a Bayesian inference model, change points are sampled and converted to a categorical distribution to compute the likelihood of the observations given the samples, as in the *independent phase mapping*. The previous model is extended by using each defined categorical region (bounded by either a change point or the edge of the search space) to define a phase region $r$. The functional property in each phase region is represented by an independent radial basis function kernel Gaussian process, with Bayesian inference sampled set of parameters $\sigma_{i,r}$: $l_{i,r}$ kernel length scale, $s_{i,r}$ kernel standard deviation, $s_i$ noise standard deviation – here it is assumed the measured noise is the same across all phase regions. Together, $GP_{i,r}$ the phase-region-bound GPs form $GP_i$ a piecewise GP. The likelihood is then computed for the observed functional property data. Bayesian inference sampling is guided by the sum of likelihoods from functional property regression and phase mapping. $P_y$ the Bayesian posterior for the functional property is approximated from the Bayesian inference samples by drawing a set of subsamples $y_{GP}$ from each $GP_i$ and computing a mean and standard deviation over the subsamples. The mean and standard deviation are then employed in a multivariate normal distribution (MVN).

$$\underline{\text{Bayesian inference step } i:}$$
$$c_i \sim Uniform(X)$$
$$l_{i,r} \sim Uniform(1, 20)$$
$$s_{i,r} \sim Uniform(1, 20)$$
$$n_i \sim Uniform(0.01, 0.1)$$
$$M_{i,s} = membership(c_i, x_s)$$
$$L_s = \sum_i \ln[Categorical(y_s(i)|M_s)]$$
$$L_f = \sum_r \sum_i \ln\left[P\left(y_f^r(i)\middle| GP_{i,r}(x_f^r, y_f^r | l_{i,r}, s_{i,r}, n_i)\right)\right]$$

$$\underline{\text{Bayesian analysis}}$$
$$P_M = \text{mean}[membership(c_i, X)]$$
$$y_{GP_i} \sim GP_i$$
$$P_y \approx \text{MVN}(\text{mean}[y_{GP}|_i], \text{std}[y_{GP}|_i])$$

## Agents: Independent Decision Making

PM agents, which are focused on learning the composition-structure relationship, select subsequent compositions based on entropy over $P_M$. FP agents, which are focused on learning composition-property relationship use Gaussian Process Upper Confidence Bounds[29] (GP-UCB) given by $\text{argmax}_x[\mu + \sigma\sqrt{\ln(Dn^2\pi^2)/3\lambda}]$. Here, μ and σ are the GP mean and standard deviation, respectively; n is the current iteration number; at each iteration a grid of D compositions is selected to search over; λ is a predefined constant, here set to 0.1.

## Agents: Joint Decision Making

For this demonstration, each FP agent $k$ takes their initial GP-UCB acquisition function $\alpha_k$ and combines it with $\bar{\alpha}_{pm}$ the mean of the acquisition functions of the PM agents through the scheduled, weighted function: $w\alpha_k + (1-w)\bar{\alpha}_{pm}$. By combining acquisition functions focused on function property optimization with identification of changepoints, the combined functional property acquisition function will have greater utility near the phase boundaries. In other words, investigating material $x$ has greater desirability if either it is associated with greater likelihood of being a change point location, or being the location of a functional property maximum. Here the weight $w$ is given by $w = \min[\max[s_{cp}], 2]/2$, where the probability of each changepoint is represented by a normal distribution $N(\mu_{cp}, s_{cp})$ where $\mu_{cp}$ is the mean and $s_{cp}$ the standard deviation, and $s_{cp}$ is the set of standard deviations for all changepoints. The value of 2 is used to provide a bound for the desired changepoint uncertainty through its standard deviation. As we have computed a probability for the location of each changepoint, this knowledge can be further exploited if one assumes functional property extrema occur at either the edge or center of a phase region using a function similar to that of [11].

## Data Repositories

Each data repository is designed as an object with a Pandas DataFrame-based database and operational functions for common tasks such as adding and updating entries.

## Discrete Events and Resources

The discrete events and resources are built using the Simpy library[30]. Measurement times and synthesis times are set to 1 time unit.

**Acknowledgements**: We thank Ichiro Takeuchi for discussions on the $Bi_{1-x}Sm_xFeO_3$ material system.

**Author Contributions:** A.G.K. made primary contributions to concept, experiments, software development, and writing, with support in all aspects by A.M.

**Declaration of Interests:** The authors declare no competing interests.

**References**:
1. Stach, E., DeCost, B., Kusne, A.G., Hattrick-Simpers, J., Brown, K.A., Reyes, K.G., Schrier, J., Billinge, S., Buonassisi, T., and Foster, I. (2021). Autonomous experimentation systems for materials development: A community perspective. Matter *4*, 2702–2726.

2. Settles, B. (2010). Active learning literature survey. University of Wisconsin, Madison *52*, 11.

3. Baker, N., Alexander, F., Bremer, T., Hagberg, A., Kevrekidis, Y., Najm, H., Parashar, M., Patra, A., Sethian, J., Wild, S., et al. (2019). Workshop Report on Basic Research Needs for Scientific Machine Learning: Core Technologies for Artificial Intelligence (USDOE Office of Science (SC), Washington, D.C. (United States)) 10.2172/1478744.

4. Gongora, A.E., Xu, B., Perry, W., Okoye, C., Riley, P., Reyes, K.G., Morgan, E.F., and Brown, K.A. (2020). A Bayesian experimental autonomous researcher for mechanical design. Science advances *6*, eaaz1708.

5. Nikolaev, P., Hooper, D., Webber, F., Rao, R., Decker, K., Krein, M., Poleski, J., Barto, R., and Maruyama, B. (2016). Autonomy in materials research: a case study in carbon nanotube growth. npj Computational Materials *2*, 16031. 10.1038/npjcompumats.2016.31.

6. Attia, P.M., Grover, A., Jin, N., Severson, K.A., Markov, T.M., Liao, Y.-H., Chen, M.H., Cheong, B., Perkins, N., and Yang, Z. (2020). Closed-loop optimization of fast-charging protocols for batteries with machine learning. Nature *578*, 397–402.

7. MacLeod, B.P., Parlane, F.G., Morrissey, T.D., Häse, F., Roch, L.M., Dettelbach, K.E., Moreira, R., Yunker, L.P., Rooney, M.B., and Deeth, J.R. (2020). Self-driving laboratory for accelerated discovery of thin-film materials. Science Advances *6*, eaaz8867.

8. Burger, B., Maffettone, P.M., Gusev, V.V., Aitchison, C.M., Bai, Y., Wang, X., Li, X., Alston, B.M., Li, B., and Clowes, R. (2020). A mobile robotic chemist. Nature *583*, 237–241.

9. Langner, S., Häse, F., Perea, J.D., Stubhan, T., Hauch, J., Roch, L.M., Heumueller, T., Aspuru-Guzik, A., and Brabec, C.J. (2020). Beyond ternary OPV: high-throughput experimentation and self-driving laboratories optimize multicomponent systems. Advanced Materials *32*, 1907801.

10. Ament, S., Amsler, M., Sutherland, D.R., Chang, M.-C., Guevarra, D., Connolly, A.B., Gregoire, J.M., Thompson, M.O., Gomes, C.P., and Dover, R.B. van (2021). Autonomous materials synthesis via hierarchical active learning of nonequilibrium phase diagrams. Science Advances. 10.1126/sciadv.abg4930.

11. Kusne, A.G., Yu, H., Wu, C., Zhang, H., Hattrick-Simpers, J., DeCost, B., Sarker, S., Oses, C., Toher, C., and Curtarolo, S. (2020). On-the-fly closed-loop materials discovery via Bayesian active learning. Nature communications *11*, 1–11.

12. McDannald, A., Frontzek, M., Savici, A.T., Doucet, M., Rodriguez, E.E., Meuse, K., Opsahl-Ong, J., Samarov, D., Takeuchi, I., and Ratcliff, W. (2022). On-the-fly autonomous control of neutron diffraction via physics-informed Bayesian active learning. Applied Physics Reviews *9*, 021408.

13. Yang, Q., Liu, Y., Chen, T., and Tong, Y. (2019). Federated machine learning: Concept and applications. ACM Transactions on Intelligent Systems and Technology (TIST) *10*, 1–19.

14. Dorri, A., Kanhere, S.S., and Jurdak, R. (2018). Multi-agent systems: A survey. Ieee Access *6*, 28573–28593.

15. Cao, K., Liu, Y., Meng, G., and Sun, Q. (2020). An Overview on Edge Computing Research. IEEE Access *8*, 85714–85728. 10.1109/ACCESS.2020.2991734.

16. Gomes, C.P., Bai, J., Xue, Y., Björck, J., Rappazzo, B., Ament, S., Bernstein, R., Kong, S., Suram, S.K., and van Dover, R.B. (2019). CRYSTAL: a multi-agent AI system for automated mapping of materials' crystal structures. MRS Communications *9*, 600–608.

17. H. Montoya, J., T. Winther, K., A. Flores, R., Bligaard, T., S. Hummelshøj, J., and Aykol, M. (2020). Autonomous intelligent agents for accelerated materials discovery. Chemical Science *11*, 8517–8532. 10.1039/D0SC01101K.

18. Rahmanian, F., Flowers, J., Guevarra, D., Richter, M., Fichtner, M., Donnely, P., Gregoire, J.M., and Stein, H.S. (2022). Enabling Modular Autonomous Feedback-Loops in Materials Science through Hierarchical Experimental Laboratory Automation and Orchestration. Advanced Materials Interfaces *9*, 2101987.

19. Godsil, C., and Royle, G.F. (2001). Algebraic graph theory (Springer Science & Business Media).

20. Graef, M.D., and McHenry, M.E. (2012). Structure of Materials: An Introduction to Crystallography, Diffraction and Symmetry 2 edition. (Cambridge University Press).

21. Lutz, M. (2013). Learning python: Powerful object-oriented programming ( O'Reilly Media, Inc.).

22. Kan, D., Pálová, L., Anbusathaiah, V., Cheng, C.J., Fujino, S., Nagarajan, V., Rabe, K.M., and Takeuchi, I. (2010). Universal Behavior and Electric-Field-Induced Structural Transition in Rare-Earth-Substituted BiFeO3. Advanced Functional Materials *20*, 1108–1115. 10.1002/adfm.200902017.

23. Koller, D., and Friedman, N. (2009). Probabilistic graphical models: principles and techniques (MIT press).

24. Tesfatsion, L. (2006). Agent-based computational economics: A constructive approach to economic theory. Handbook of computational economics *2*, 831–880.

25. Gao, M., Wang, Z., Wang, K., Liu, C., and Tang, S. (2022). Forecasting elections with agent-based modeling: Two live experiments. PLOS ONE *17*, e0270194. 10.1371/journal.pone.0270194.

26. Bingham, E., Chen, J.P., Jankowiak, M., Obermeyer, F., Pradhan, N., Karaletsos, T., Singh, R., Szerlip, P., Horsfall, P., and Goodman, N.D. (2019). Pyro: Deep universal probabilistic programming. The Journal of Machine Learning Research *20*, 973–978.

27. Matthews, A.G. de G., Van Der Wilk, M., Nickson, T., Fujii, K., Boukouvalas, A., León-Villagrá, P., Ghahramani, Z., and Hensman, J. (2017). GPflow: A Gaussian Process Library using TensorFlow. J. Mach. Learn. Res. *18*, 1–6.

28. Zhu, X., Lafferty, J., and Ghahramani, Z. (2003). Combining active learning and semi-supervised learning using gaussian fields and harmonic functions. In ICML 2003 workshop on the continuum from labeled to unlabeled data in machine learning and data mining.

29. Shahriari, B., Swersky, K., Wang, Z., Adams, R.P., and de Freitas, N. (2016). Taking the Human Out of the Loop: A Review of Bayesian Optimization. Proceedings of the IEEE *104*, 148–175. 10.1109/JPROC.2015.2494218.

30. Matloff, N. (2008). Introduction to discrete-event simulation and the simpy language. Davis, CA. Dept of Computer Science. University of California at Davis. Retrieved on August *2*, 1–33.

**Figures:**

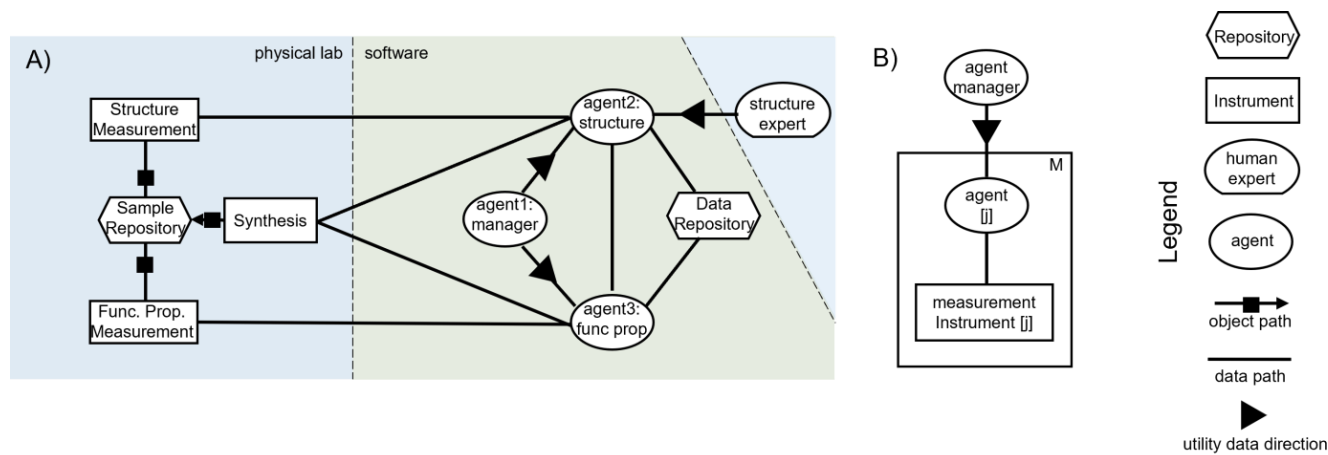

Figure 1. Visual language for presenting framework implementations. A) implementation diagram including representation of the physical lab (blue region) and the software (green region). The legend can be found to the right of the figure. Connections between objects indicate information or sample flow paths. The physical lab consists of synthesis, structure measurement, and functional property instruments, a sample repository, and a human structure expert. Edges marked with '■' indicate sample paths. The software consists of agents and a data repository. Edges indicate paths for data transfer and the icon ► indicates additional sharing of acquisition function with directionality pointing from leader to follower. If both directions are present, each agent incorporates the others acquisition function data. B) The common 'plate' representation [22] can be used to indicate a number (here, M) of identical units.

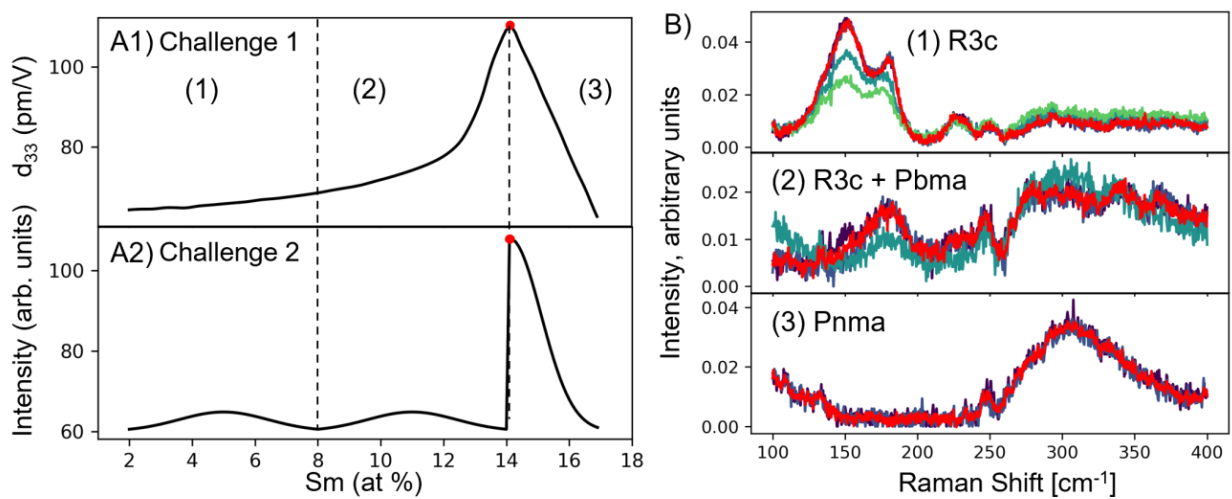

Figure 2. The two materials optimization challenges. The materials data challenge is drawn from the perovskite oxide material system as characterized by [22] for piezoelectric response and structure. Data for samples near the $Bi_{1-x}Sm_xFeO_3$ edge of this composition space is shown. A1) The functional property challenge is to identify the maximum value of the piezoelectric coefficient $d_{33}$ (pm/V) (maximum indicated with red dot) and the associated composition. This parameter is strongly dependent on the phase diagram as indicated by the maximum located near the boundary between phases (2) and (3). The maximum occurs with a single peak which extends the entire shown composition range. A2) A more complex synthetic example where the target functional property peaks near the second phase boundary and is characterized by a narrow peak. B) Raman spectra for samples in the phase regions (1-3) are shown along with the lattice space groups present.

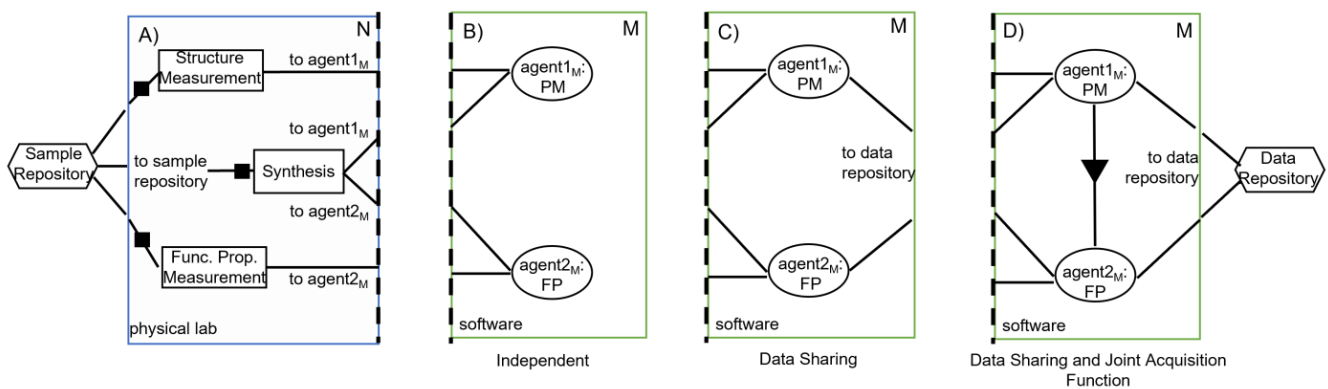

Figure 3. Diagrams for portions of facility units. Here we use the plate notation with M multiples. A) The physical lab portion of the facility unit consists of a sample repository and the following instruments: sample synthesis, Raman-based structure measurement, and $d_{33}$ functional property measurement. One sample repository is used for the M repeated objects. B) The software portion of the facility unit, containing two independent agents, agent$1_M$:PM focused on synthesis-structure relationship (i.e., phase map) and agent$2_M$:FP focused on the synthesis-property relationship. C) an alternative software portion where agents share sample data through the data repository. All agents share the same data repository. D) an alternative software portion where agents share sample data and acquisition functions. All agents share the same data repository.

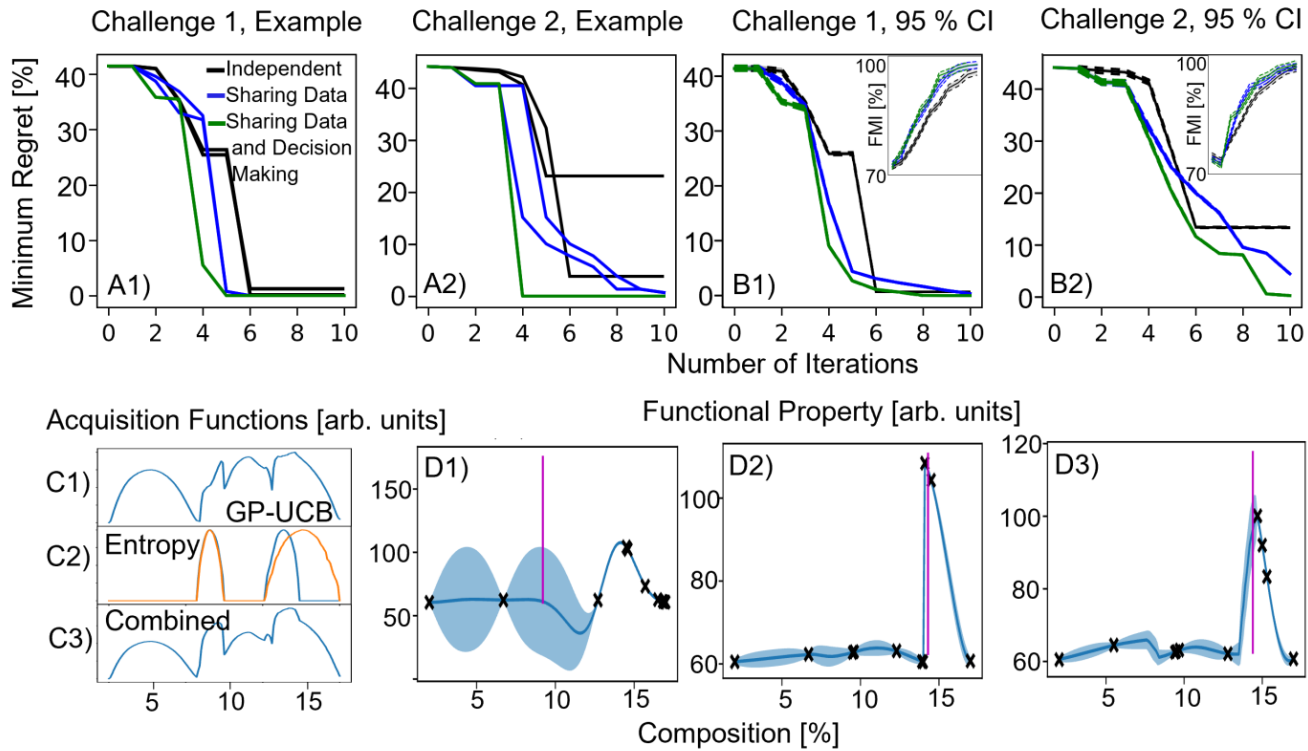

Figure 4. Materials optimization performance for different agent architectures. Materials optimization for A1) one example run of challenge 1 and, A2) challenge 2. Performance is shown using % minimum regret. B1 and B2 show the 95 % confidence interval for the performance mean over ten runs for each challenge. The insets present the percent Fowlkes Mallow score for the corresponding phase mapping. Example of combining acquisition functions after 5 data points are collected: (C1) Before combination, using Gaussian Process Upper Confidence Bounds for agent $FP_1$, (C2) entropy acquisition functions for 2 phase mapping agents ($PM_1$, $PM_2$), (C3) combined acquisition function with increased desirability for compositions that impact both phase mapping and functional property optimization. D) Example functional property predictions for Challenge 2 after 10 iterations: of (D1) FP1 in the Independent architecture from Fig. 3B, where knowledge of the phase diagram is not employed and a single functional behavior is assumed across the composition search space, (D2) of agent FP1 for the Data Sharing

coregionalization architecture from Figure 3C and (D3) of agent FP1 for the Data Sharing and Joint Decision Making coregionalization architecture with acquisition sharing from Fig. 3D. For these latter two, knowledge of phase boundaries allows for different functional property behavior for each prediction phase region.